%% file: NuPhys_proceedings.tex
\documentclass[12pt]{article}
\usepackage{graphicx}
\usepackage[justification=centering]{caption}

\newcommand\pubnumber{NuPhys2018-Paton}
\newcommand\pubdate{\today}

\def\oxford{Department of Physics\\ University of Oxford, Oxford, UK}

\def\Title#1{\begin{center} {\Large #1 } \end{center}}
\def\Author#1{\begin{center}{ \sc #1} \end{center}}
\def\Address#1{\begin{center}{ \it #1} \end{center}}

\newcommand\pubblock{\rightline{\begin{tabular}{l} \pubnumber\\
         \pubdate  \end{tabular}}}
\newenvironment{Abstract}{\begin{quotation}  }{\end{quotation}}
\newenvironment{Presented}{\begin{quotation} \begin{center} 
             PRESENTED AT\end{center}\bigskip 
      \begin{center}\begin{large}}{\end{large}\end{center} \end{quotation}}

\input econfmacros.tex

\begin{document}
\begin{titlepage}
\pubblock

\vfill
\Title{Neutrinoless Double Beta Decay in the SNO+ Experiment}
\vfill
\Author{ Josephine Paton\\On behalf of the SNO+ Collaboration}
\Address{\oxford}
\vfill
\begin{Abstract}

SNO+ is a large multipurpose neutrino detector situated 2km underground at SNOLAB in Sudbury, Canada~\cite{SNOplus2016}. It reuses the structure of the SNO experiment with numerous infrastructure upgrades and with heavy water replaced by ultra-pure liquid scintillator. The detector will be loaded with~0.5\% natural tellurium in order to search for neutrinoless double beta decay ($0\nu\beta\beta$). The expected sensitivity after 5 years of data taking is T$^{0\nu} _{1/2} > 1.9 \times 10^{26}$~years (90\% CL)~\cite{Fischer2018}. A future increase in loading could achieve a sensitivity of ${\sim}10^{27}$~years. 
 
\end{Abstract}
\vfill
\begin{Presented}
NuPhys2018, Prospects in Neutrino Physics\\

Cavendish Conference Centre, London, UK, December 19--21, 2018
\end{Presented}
\vfill
\end{titlepage}
\def\thefootnote{\fnsymbol{footnote}}
\setcounter{footnote}{0}

\section{Introduction to SNO+}
The SNO+ detector contains a 12~m diameter acrylic vessel filled surrounded by a geodesic PMT support structure (PSUP) of ${\sim}18 $~m radius. Mounted on the PSUP are ${\sim} 9,300$ inward facing PMTs for 50\% effective coverage. The detector is positioned inside a 40~m tall cavern filled with ultra pure water (UPW) for shielding~\cite{SNOplus2016}.

There are three main stages of the SNO+ experiment, based on the detection material inside the acrylic vessel. These are UPW, scintillator and $^{130}$Te loaded scintillator. The detector will initially be loaded with 0.5\%  $^{130}$Te by mass. Scintillator filling is currently in progress. Higher loading concentrations would be possible as a further upgrade.

\section{Tellurium Loaded Scintillator Cocktail}

In the double beta decay ($\beta\beta$) phase of SNO+, the acrylic vessel will be filled with Tellurium loaded scintillator in order to search for $0\nu\beta\beta$. $^{130}$Te is a double beta decay isotope with a Q-value of 2.527~MeV~\cite{Ouellet2018}. In $^{130}$Te, the two-neutrino double beta decay ($2\nu\beta\beta$) rate relative to $0\nu\beta\beta$ is notably smaller than in other isotopes (excluding $^{136}$Xe), leading to a comparatively low background. Also, the $2\nu\beta\beta$ lifetime is large ($7.9 \times 10^{20}$~years)~\cite{Grant2018}, leading to a low absolute rate of $2\nu\beta\beta$ events. $^{130}$Te has a high natural abundance (34\%) and can be loaded with a good optical transparency in scintillator. 

\subsection{Scintillator Cocktail}
\begin{figure}[!h]
\centering
\includegraphics[width=3.7in]{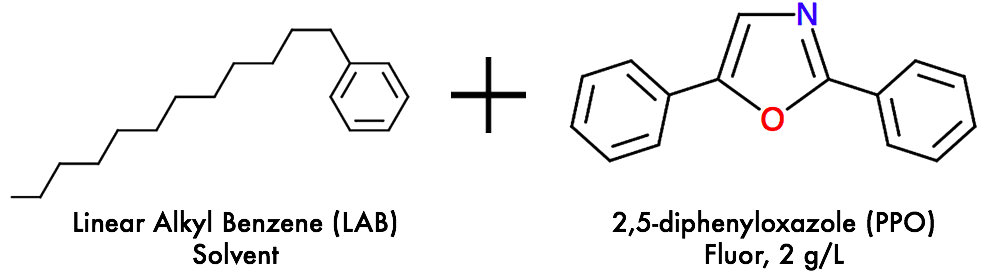}
\caption{Composition of the scintillator cocktail~\cite{Manecki2018}}
\label{fig:Scint}
\end{figure}

The scintillator cocktail has two main components: linear alkyl benzene (LAB) as a solvent and 2,5-diphenyloxazole (PPO) as a fluor. The PPO has a concentration of 2~g/L. These constituents can be found in Figure \ref{fig:Scint}.

Included in the cocktail is N,N-dimethyldodecylamine (DDA). This improves the light yield of the scintillator and stabilises the full cocktail against humidity exposure.

\subsection{Tellurium Loading}

\begin{figure}[!h]
\centering
\includegraphics[width=5in]{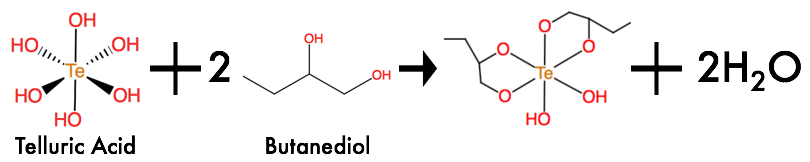}
\caption{Initial stages of compound formation~\cite{Manecki2018}}
\label{fig:TeDiol}
\end{figure}

In order to load the $^{130}$Te into the scintillator, an organo-metallic compound is formed from telluric acid and butanediol~\cite{Grant2018}. The initial stages of this process are depicted in Figure \ref{fig:TeDiol}. This compound is then loaded into the scintillator. This method allows the  optical transparency of the scintillator to be conserved. The total light yield of the full cocktail is expected to be 400~NHits/MeV~\cite{Grant2018}.

\section{Backgrounds and Sensitivities}
 
\begin{figure}[h!]
\centering
\includegraphics[width=4.8in]{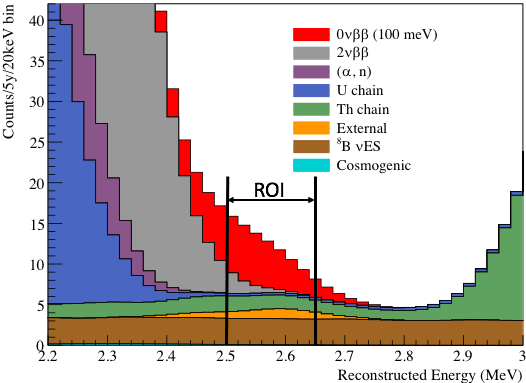}
\caption{Energy spectrum of $0\nu\beta\beta$ signal (red) and background\cite{Backgrounds2018} \hspace{\textwidth}Region of interest: 2.49 - 2.65 MeV}
\label{fig:spectrum}
\end{figure}

The expected energy spectrum of both the $0\nu\beta\beta$ signal and the backgrounds within the fiducial volume are shown in Figure \ref{fig:spectrum}. The fiducial volume extends to a radius of 3.3~m in order to shield from external backgrounds. Using this spectrum, the region of interest (ROI) is selected to be 2.49-2.65~MeV~\cite{Backgrounds2018}, containing the bulk of the $0\nu\beta\beta$ signal while also minimising the background.

\begin{figure}[h!]
\centering
\includegraphics[width=4.8in]{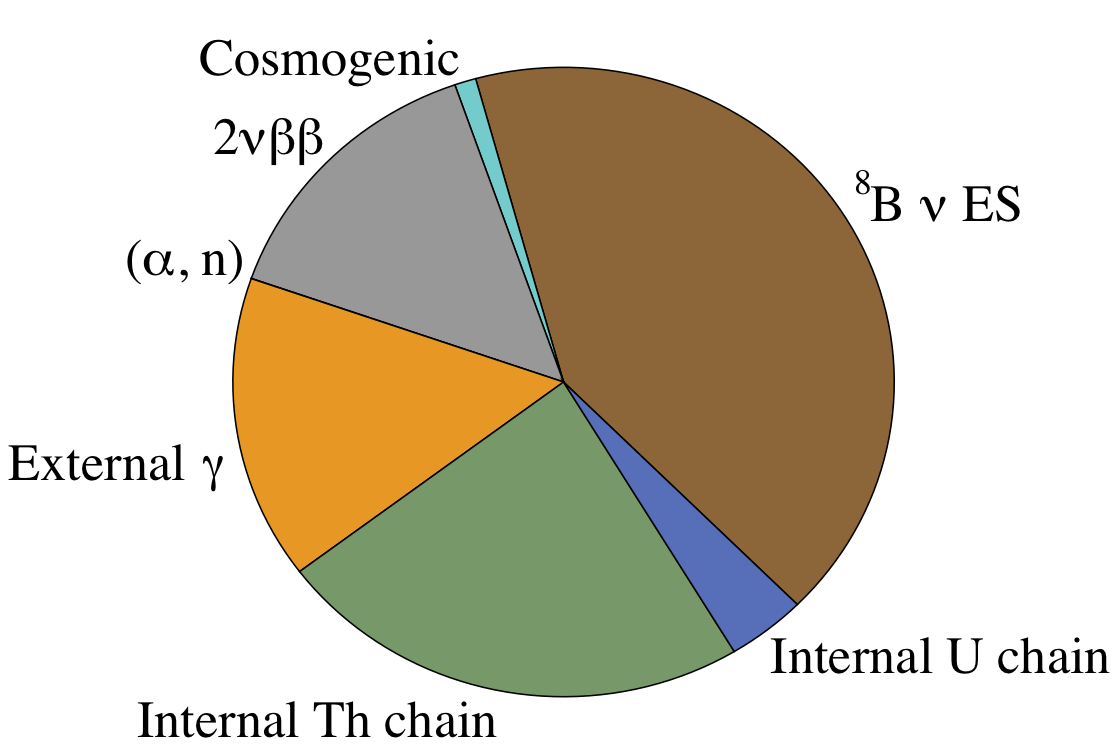}
\caption{Background counts within ROI in first year of data taking\cite{Backgrounds2018} \hspace{\textwidth}Total counts: 12.4}
\label{fig:pie}
\end{figure}

An evaluation of background sources can be seen in Figure \ref{fig:pie}. These backgrounds are predictions for the first year of data taking within the region of interest. It can be seen that the dominant background is elastic scattering from $^8$B solar neutrinos; however, this background is constant across the energy range. 

\begin{figure}[h!]
\centering
\includegraphics[width=4.8in]{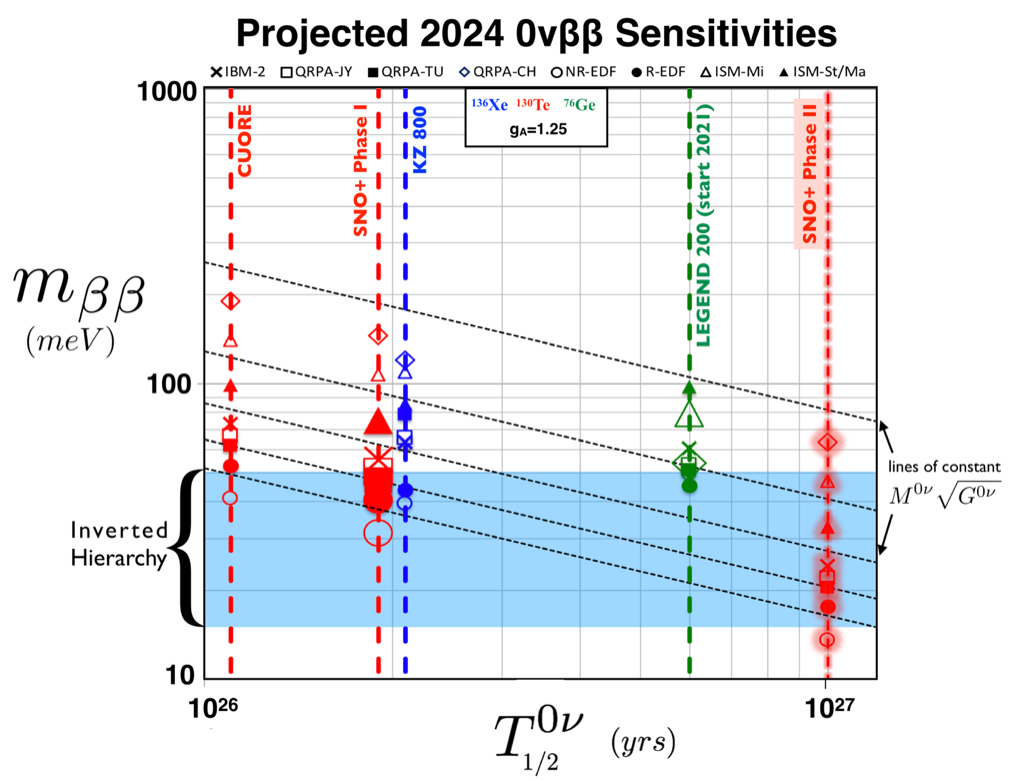}
\caption{Projections of $0\nu\beta\beta$ experiments, including SNO+ Phase I and II\cite{Manecki2018}}
\label{fig:project}
\end{figure}

Figure \ref{fig:project} shows the projected sensitivities for several $0\nu\beta\beta$ experiments. SNO+ Phase I has projected sensitivity of T$^{0\nu}_{1/2} > 1.9 \times 10^{26}$~years (90\% CL)~\cite{Backgrounds2018}. This corresponds to a limit of m$_{\beta\beta} < 41 -99$~meV~\cite{Backgrounds2018}, with the range defined by the models listed at the top of Figure \ref{fig:project}. This mass range probes the inverted hierarchy.

Also included in the figure is the projected sensitivity of a SNO+ Phase II experiment. It is expected that a second phase of the SNO+ experiment could reach a T$^{0\nu}_{1/2}$ limit of at least $10^{27}$~years. This could be achieved with increased $^{130}$Te loading. This new limit would correspond to a mass range that probes beyond the inverted hierarchy. 

 \pagebreak

\bibliographystyle{unsrt}
\bibliography{NuPhysBib.bib}{}

\end{document}

%% file: econfmacros.tex
\def\beq{\begin{equation}}
\def\eeq#1{\label{#1}\end{equation}}
\def\eeqn{\end{equation}}

\def\beqa{\begin{eqnarray}}
\def\eeqa#1{\label{#1}\end{eqnarray}}
\def\eeqan{\end{eqnarray}}

\let\bar=\overbar

\def\Dslash{\not{\hbox{\kern-4pt $D$}}}
\def\dslash{\not{\hbox{\kern-2pt $\del$}}}

\def\msb{{\bar{\ssstyle M \kern -1pt S}}}

